\documentclass[%
 reprint,
superscriptaddress,
 amsmath,amssymb,
 aps,
floatfix,
showkeys
]{revtex4-2}

\usepackage{graphicx}
\usepackage{dcolumn}
\usepackage{bm}
\usepackage[utf8]{inputenc}
\usepackage{tabularx}
\usepackage[dvipsnames]{xcolor}
\usepackage[euler]{textgreek}
\usepackage{upgreek}
\usepackage{derivative}
\usepackage{amsmath}
\usepackage{physics}
\usepackage[caption=false]{subfig}
\usepackage[version=4]{mhchem}
\usepackage[separate-uncertainty=true, free-standing-units=true, binary-units=true]{siunitx}
\usepackage{amssymb}
\usepackage[colorlinks=true,linkcolor=blue!50!black,urlcolor=blue!50!black,citecolor=blue!50!black]{hyperref}
\usepackage{cleveref}

\begin{document}

\preprint{APS}

\newcommand{\mpi}{\affiliation{Max-Planck-Institut f\"ur Physik, 85748 Garching, Germany}}
\newcommand{\coimbra}{\affiliation{Also at: LIBPhys, Departamento de Fisica, Universidade de Coimbra, P3004 516 Coimbra, Portugal}}
\newcommand{\hephy}{\affiliation{Institut f\"ur Hochenergiephysik der \"Osterreichischen Akademie der Wissenschaften, 1010 Wien, Austria}}
\newcommand{\ati}{\affiliation{Atominstitut, Technische Universit\"at Wien, 1020 Wien, Austria}}
\newcommand{\tum}{\affiliation{Physik-Department, TUM School of Natural Sciences, Technische Universit\"at M\"unchen, D-85747 Garching, Germany}}
\newcommand{\tuebingen}{\affiliation{Eberhard-Karls-Universit\"at T\"ubingen, 72076 T\"ubingen, Germany}} 
\newcommand{\bratislava}{\affiliation{Comenius University, Faculty of Mathematics, Physics and Informatics, 84248 Bratislava, Slovakia}}
\newcommand{\kip}{\affiliation{Kirchhoff-Institute for Physics, Heidelberg University, 69120 Heidelberg, Germany}}
\newcommand{\iap}{\affiliation{Institute for Astroparticle Physics, Karlsruhe Institute of Technology, 76128 Karlsruhe, Germany}}

\newcommand{\oxford}{\affiliation{Department of Physics, University of Oxford, Oxford OX1 3RH, United Kingdom}}
\newcommand{\wmi}{\affiliation{Also at: Walther-Mei\ss ner-Institut f\"ur Tieftemperaturforschung, 85748 Garching, Germany}}
\newcommand{\lngs}{\affiliation{INFN, Laboratori Nazionali del Gran Sasso, 67010 Assergi, Italy}}
\newcommand{\cassino}{\affiliation{Also at: Dipartimento di Ingegneria Civile e Meccanica, Università degli Studi di Cassino e del Lazio Meridionale, 03043 Cassino, Italy}}
\newcommand{\usp}{\affiliation{Also at: Instituto de F\'{i}sica, Universidade de S$\tilde{a}$o Paulo, S$\tilde{a}$o Paulo 05508-090, Brazil}}
\newcommand{\umb}{\affiliation{Also at: Dipartimento di Fisica, Universit\`a di Milano Bicocca, Milano, 20126, Italy}}

\newcommand{\prague}{\affiliation{Institute of Experimental and Applied Physics, Czech Technical University in Prague, 110 00 Prague 1, Czech Republic}}
\newcommand{\jgu}{\affiliation{Present address: Johannes Gutenberg-Universit\"at Mainz, Institut f\"ur Physik, 55128 Mainz, Germany}}

\mpi
\hephy
\ati
\kip
\bratislava
\lngs
\tuebingen
\tum
\oxford
\iap

\coimbra
\wmi
\usp
\cassino
\umb

\prague
\jgu

\author{G.~Angloher}
  \mpi

\author{S.~Banik}
    \email[Corresponding author: ]{samir.banik@oeaw.ac.at}
  \hephy
  \ati

\author{A.~Bento}
  \mpi
  \coimbra 

\author{A.~Bertolini}
  \kip

\author{R.~Breier}
  \bratislava

\author{C.~Bucci}
  \lngs

\author{L.~Burmeister}
    \kip

\author{F.~Casadei}
    \mpi

\author{E.~Cipelli}
    \mpi

\author{J.~Burkhart}
    \email[Corresponding author: ]{jens.burkhart@oeaw.ac.at}
  \hephy

\author{L.~Canonica}
  \mpi 

\author{J.~Dohm}
    \tuebingen
    
\author{F.~Dominsky}
    \mpi

\author{S.~Di~Lorenzo}
  \mpi
  \lngs

\author{L.~Einfalt}
  \hephy
  \ati
  
\author{A.~Erb}
  \tum
  \wmi

\author{E.~Fascione}
    \kip
  
\author{F.~v.~Feilitzsch}
  \tum  
  
 \author{S.~Fichtinger}
  \hephy
 
\author{D.~Fuchs}
    \email[Corresponding author: ]{dominik.fuchs@oeaw.ac.at}
  \mpi
  \hephy
  \ati
  
 \author{V.M.~Ghete}
  \hephy 

\author{P.~Gorla}
  \lngs 

\author{P.V.~Guillaumon}
  \mpi
  \usp

\author{D.~Hauff}
  \mpi 

\author{M.~Ješkovsk\'y}
  \bratislava

\author{J.~Jochum}
  \tuebingen 

\author{M.~Kaznacheeva}
  \tum
  
\author{H.~Kluck}
    \email[Corresponding author: ]{holger.kluck@oeaw.ac.at}
  \hephy

\author{H.~Kraus}
  \oxford

\author{B.~von~Krosigk}
    \kip
    \iap

\author{A.~Langenk\"amper}
  \mpi

\author{M.~Mancuso}
  \mpi

\author{B.~Mauri}
    \mpi
  
\author{V.~Mokina}
  \hephy

\author{C.~Moore}
    \mpi

\author{P.~Murali}
    \kip

\author{M.~Olmi}
  \lngs
  
\author{T.~Ortmann}
  \tum

\author{C.~Pagliarone}
  \lngs 
  \cassino

\author{L.~Pattavina}
  \lngs
  \umb

\author{F.~Petricca}
  \mpi 

\author{W.~Potzel}
  \tum 

\author{P.~Povinec}
  \bratislava

\author{F.~Pr\"obst}
  \mpi

\author{F.~Pucci}
  \lngs
  
\author{F.~Reindl}
  \hephy
  \ati

\author{J.~Rothe}
  \tum
  
\author{K.~Sch\"affner}
  \mpi

\author{J.~Schieck}
  \hephy
  \ati 

\author{S.~Sch\"onert}
  \tum 
  
\author{C.~Schwertner}
  \hephy
  \ati

\author{M.~Stahlberg}
  \mpi

\author{L.~Stodolsky}
  \mpi 

\author{C.~Strandhagen}
  \tuebingen

\author{R.~Strauss}
  \tum

\author{I.~Usherov}
  \tuebingen 

\author{D.~Valdenaire}
    \hephy
    \ati

\author{M.~Zanirato}
    \mpi

\author{V.~Zema}
  \mpi
  \hephy

\collaboration{CRESST Collaboration}
    \noaffiliation

\author{M.~Macko}
    \prague
    
\author{V.~Palušová}
    \prague
    \jgu

\title{Observation of a low energy nuclear recoil peak in the neutron calibration data of an Al$_{2}$O$_{3}$ crystal in CRESST-III}

\begin{abstract}
The current generation of cryogenic solid state detectors used in direct dark matter and CE\textnu NS searches typically reach energy thresholds of $\mathcal{O}$(10)$\,$eV for nuclear recoils. For a reliable calibration in this energy regime a method has been proposed, providing mono-energetic nuclear recoils at low energies $\sim\,$100$\,$eV$\,$-$\,$1$\,$keV. In this work we report on the observation of a peak at (1113.6$^{+6.5}_{-6.5}$)$\,$eV in the data of an Al$_{2}$O$_{3}$ crystal in CRESST-III, which was irradiated with neutrons from an AmBe calibration source. We attribute this mono-energetic peak to the radiative capture of thermal neutrons on $^{27}$Al and the subsequent de-excitation via single $\gamma$-emission. We compare the measured results with the outcome of Geant4 simulations and investigate the possibility to make use of this effect for the energy calibration of Al$_{2}$O$_{3}$ detectors at low energies. We further investigate the possibility of a shift in the expected energy scale of this effect caused by the creation of defects in the target crystal.
\end{abstract}

\keywords{Cryogenic detectors, Dark matter, CE\textnu NS}
\maketitle



\section{Introduction}

Many experiments employing cryogenic solid state detectors for the search of low-mass dark matter (DM) or Coherent Elastic Neutrino-Nucleus Scattering (CE\textnu NS) have optimized their detectors to be sensitive to nuclear recoils at energies of $\mathcal{O}$(10)$\,$eV \cite{PhysRevD.96.022009, PhysRevD.100.102002, Angloher2024, PhysRevD.99.082003, PhysRevD.106.062004, PhysRevLett.127.061801, PhysRevLett.121.051301, PhysRevD.105.112006} or even below \cite{PhysRevD.107.122003, AngloherVUV}. The energy calibration of these detectors is usually done via X-ray sources with energies of several keV, requiring an extrapolation to the low-energy regime. Additionally, the accuracy of the energy calibration at low energies can be impacted by differences between electromagnetic interactions (as by the X-ray source) and the sought for nuclear recoils, e.g. caused by energy loss due to defect creation in the crystals \cite{kadribasic2020crystal, PhysRevD.106.083009, PhysRevD.106.063012, SoumSidikov2024, Nordlund2024}. Thus, a novel calibration method is required, which ideally provides mono-energetic peaks in the low-energy nuclear recoil spectrum.

The CRAB (Calibrated nuclear Recoils for Accurate Bolometry) collaboration proposed to use a new technique, based on the radiative capture of thermal neutrons, followed by a subsequent decay via single $\gamma$-emission, leading to a nuclear recoil in a $\sim\,$100$\,$eV$\,$-$\,$1$\,$keV range \cite{Wagner2022, Thulliez_2021}. They consider detector crystals containing different suitable isotopes of W and Ge. Simulations on these materials show that clear peaks in the nuclear recoil spectrum are expected when they are exposed to a flux of thermal neutrons. First experimental proof on a CaWO$_{4}$ crystal was published by the CRAB and NUCLEUS collaborations in Ref.~\cite{PhysRevLett.130.211802} and by the CRESST collaboration in Ref.~\cite{PhysRevD.108.022005}. In this work we show that this effect is also observable in sapphire (Al$_{2}$O$_{3}$) crystals, due to the aluminum isotope $^{27}$Al having suitable properties. The energy spectrum of a sapphire crystal, obtained during a calibration measurement with an AmBe neutron source contains a peak at (1113.6$^{+6.5}_{-6.5}$)$\,$eV, which arises from nuclear recoils resulting from the emission of prompt $\gamma$ photons during the de-excitation of $^{28}$Al formed subsequent to the capture of a thermal neutron in $^{27}$Al. This peak can be used for low energy calibration, which removes the need to put other radioactive sources in the vicinity of the detector. This is advantageous, since low energetic radioactive sources close to the detector inevitably lead to an increased background and cannot be easily removed without interrupting the measurement.

Using the calibration of an $^{55}$Fe X-ray source, the fitted position of the peak is about (30.4$\,\pm\,$6.5)$\,$eV lower than expected. We investigate the possibility of this shift in energy being the result of an energy loss due to defect creation in the crystal and estimate the statistical significance of the shift as a function of the uncertainty on our energy calibration. The results show a strong dependence on the uncertainty, which is not directly accessible in the CRESST experimental setup. Depending on the value of the uncertainty, our results range from a 4$\,\sigma$ observation of a shift of the peak down to a significance of only 1.3$\,\sigma$ and are therefore by themselves not conclusive. First experimental hints of an energy loss due to defect creation were found by the SuperCDMS Collaboration in two Ge detectors \cite{10.1063/1.5041457}, in which the endpoint of the spectrum of $^{206}$Pb recoils at $\sim$100$\,$keV, originating from a $^{210}$Pb source, was measured at a $\sim$6$\,\%$ lower energy than what was expected from simulations. However the stated p-values of the measured shift in both detectors was not significant (0.08 and 0.31, respectively). A clear measurement of such a shift would be the first experimental proof of this predicted effect, which has far-reaching implications on DM and CE\textnu NS experiments. For once, this energy loss has a strong influence on the expected shape of nuclear recoil spectra, studied e.g. in Ref.~\cite{kadribasic2020crystal} for various materials (C, Ge, Si) and in Ref.~\cite{PhysRevD.106.063012} for the same materials and additional materials (WC, SiC, W), as well as specifically for CaWO$_{4}$ in Ref.~\cite{SoumSidikov2024}. Secondly, the release of the energy stored by defects in the crystal could be a potential origin of the so-called low energy excess (LEE), as suggested by the authors of Refs.~\cite{PhysRevD.106.083009, Nordlund2024}. The LEE is an unexpected rise of the event rate that exceeds expected backgrounds. A comprehensive summary on the LEE can be found in Refs.~\cite{10.21468/SciPostPhysProc.9.001, Baxter:2025odk} and a detailed study of the LEE within CRESST in Ref.~\cite{10.21468/SciPostPhysProc.12.013}. This background has an immense impact on the improvement of the sensitivity of cryogenic solid state detectors and has therefore been one of the main subjects of studies within many collaborations in recent years.

We first give a brief summary of the effect leading to the presence of the peak in Sec.~\ref{theory}, followed by a short description of the experimental setup and the detector design in Sec.~\ref{Setup}. The results of simulations of the expected signal in an Al$_{2}$O$_{3}$ detector are given in Sec.~\ref{simulations}. In Sec.~\ref{Analysis} we describe our data analysis, including the data processing and calibration as well as our likelihood framework and the description of our fit function. We show the results of this work in Sec.~\ref{Results} and end with a conclusion in Sec.~\ref{Conclusion}.

\section{Radiative neutron capture on Aluminum-27} \label{theory}
The CRAB Collaboration proposed a novel technique for an accurate calibration of nuclear recoils on CaWO$_{4}$ at the energy scale of $\sim\,$100$\,$eV$\,$-$\,$1$\,$keV in Ref.~\cite{Wagner2022}. A detailed description of the process can be found in Ref.~\cite{Thulliez_2021}. The basic principle is the capture of a thermal neutron by the nucleus, which creates a compound nucleus in an excited state. The de-excitation of this nucleus typically happens via the emission of $\gamma$-rays and conversion electrons, mostly in the form of cascades. In some cases, the nucleus de-excites directly to the ground state by emitting only a single $\gamma$ that carries the entire energy of this process. Depending on the target nucleus, these energies are usually in the order of several MeV. These de-exitations under emission of a single $\gamma$ with energy $E_{\gamma}$ lead to a monoenergetic peak in the nuclear recoil spectrum with an energy of $E_{\mathrm{R}} = E^{2}_{\gamma}/(2\cdot M_{\mathrm{N}} c^{2})$, where $M_{\mathrm{N}}$ is the mass of the compound nucleus. The energies of the recoiling nuclei $E_{\mathrm{R}}$ are typically in the order of $\sim\,$100$\,$eV$\,$-$\,$1$\,$keV. \\

The target nuclei need to fulfill certain criteria to make this technique useful for a low energy calibration:

\begin{itemize}
	\item A high natural abundance of the isotope in question, $Y_{\mathrm{ab}}$
	\item A high cross section for the capture of thermal neutrons, $\sigma_{\mathrm{n},\gamma}$
	\item A high branching ratio for single-$\gamma$ transitions, $I^{s}_{\gamma}$
\end{itemize}

The aluminum isotope $^{27}$Al in sapphire crystals has appropriate properties for this technique. The key parameters are listed in Tab.~\ref{tab:Al27_properties}.

\begin{table*}[!htb]
\centering
	\caption{\label{tab:Al27_properties}Properties of $^{27}$Al. A high natural abundance, high enough thermal neutron capture cross section and a high branching ratio for single-$\gamma$ de-excitation make it a suitable isotope to observe monoenergetic nuclear recoils caused by thermal neutron capture. All numbers are taken from Ref.~\cite{articleEGAF, BROWN20181, SHAMSUZZOHABASUNIA20131189} except for $E_{\mathrm{R}}$, which is calculated from the Q-value.}
	\newcolumntype{C}{>{\centering\arraybackslash}X}
	\setlength\extrarowheight{3pt}
	\noindent
	\begin{tabularx}{\textwidth}{ C C C C C C }
	\hline          
        Isotope & $Y_{ab}$ & $\sigma_{\mathrm{n},\gamma}$ & $I^{s}_{\gamma}$ & Q-value & $E_{\mathrm{R}}$ \\ \hline
        $^{27}$Al & 100$\,$\% & 0.23$\,$barn & 26.81$\,$\% & 7724$\,$keV & 1144$\,$eV \\ \hline
      \end{tabularx}
\end{table*}

\section{Experimental setup and detector design} \label{Setup}
\begin{figure}[!b]
\centering
\includegraphics[width=0.5\textwidth]{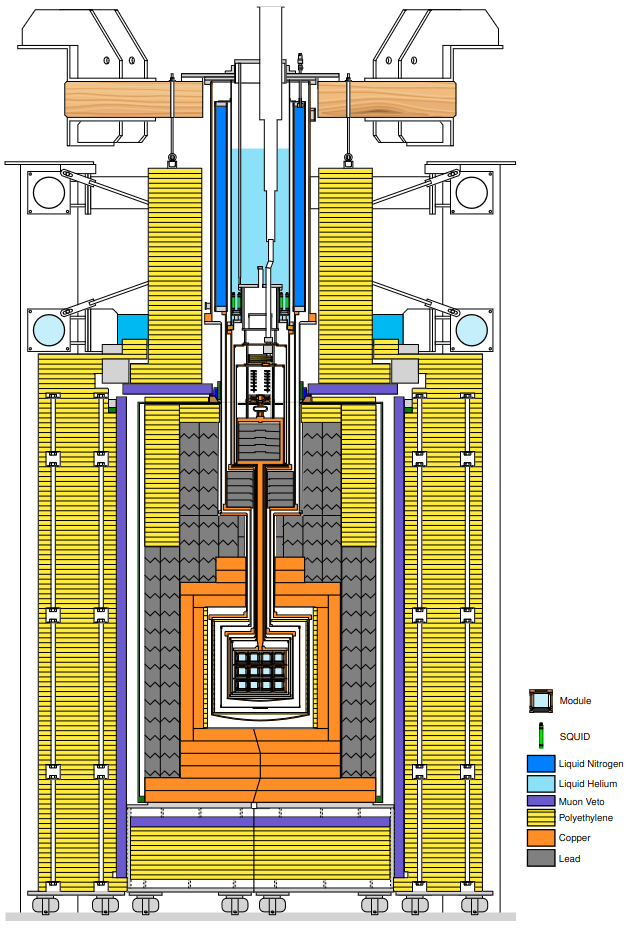}
\caption{\label{fig:Setup} Schematic cross section of the CRESST setup \cite{sdilorenzo2020}, showing the structure of the cryostat and the various layers of shielding. The detector modules are located in the center.}
\end{figure}

The CRESST-III (Cryogenic Rare Event Search with Superconducting Thermometers) experiment is located in the underground facility at the Laboratori Nazionali del Gran Sasso (LNGS) in Italy.
The shielding of the experimental setup consist of concentric layers of various materials. The entire experimental setup is shown in Fig.~\ref{fig:Setup}. \\

On the outside, a layer of polyethylene (PE) with a thickness of 40$\,$cm is moderating neutrons. Inside the PE layer, the experimental volume is shielded against radioactive backgrounds by 20$\,$cm of radiopure lead. In order to shield against remaining intrinsic radioactivity of the lead, an additional layer of 14$\,$cm of copper is used inside of the lead shield. Copper with a high radiopurity is also used for all support structures in the direct vicinity of the detectors. The innermost layer surrounding the detectors is another layer of PE, protecting against neutrons originating from the lead or copper shields. More details about the experimental setup can be found in Ref.~\cite{ANGLOHER2009270}.

The detectors are located in the center of the experimental setup, where they are cooled by a commercial $^{3}$He/$^{4}$He-dilution refrigerator, connected from above. The detector module we are discussing in this work consist of a (20$\,$x$\,$20$\,$x$\,$10)$\,$mm$^{3}$ Al$_{2}$O$_{3}$ main absorber crystal with a mass of $\sim \,$16$\,$g and a (20$\,$x$\,$20$\,$x$\,$0.4)$\,$mm$^3$ wafer detector, made of silicon-on-sapphire (SOS). Both crystals are held by copper sticks and are kept in a bare copper housing. Each detector is equipped with a tungsten transition edge sensor (W-TES), operated at a stable temperature of around 15$\,$mK in the superconducting transition. The W-TES films are stabilized via ohmic heating resistors on the crystals. We periodically inject electric pulses at different energies into these resistors, which are used in the calibration of the time dependent response of the detectors. The energies of the injected heater pulses are chosen to cover the entire dynamic range of the TES. This includes energies that are large enough to drive the detector out of its transition into the normal conducting phase, leading to saturated pulses.
The signals of the detectors are read out by SQUID-based electronics and saved by the data acquisition system \cite{Angloher2012}. The wafer detector is not relevant for the results in this work and is therefore not further mentioned. For the energy calibration, we use a low-activity $^{55}$Fe source. During the neutron calibration measurement campaign in September 2021, which lasted for about 40 days, we installed a strong AmBe source with an output of about 2000 neutrons per second inside the CRESST muon veto system, shown in purple in Fig.~\ref{fig:Setup}, but outside the lead and copper shieldings, shown in gray and orange in Fig.~\ref{fig:Setup}, at a distance of about 75$\,$cm to the detectors. The neutron irradiation does not have an impact on the performance and the radioactive contamination of the detectors.

\section{Simulations} \label{simulations}
\begin{figure}[h]
	\centering
	\includegraphics[width=0.5\textwidth]{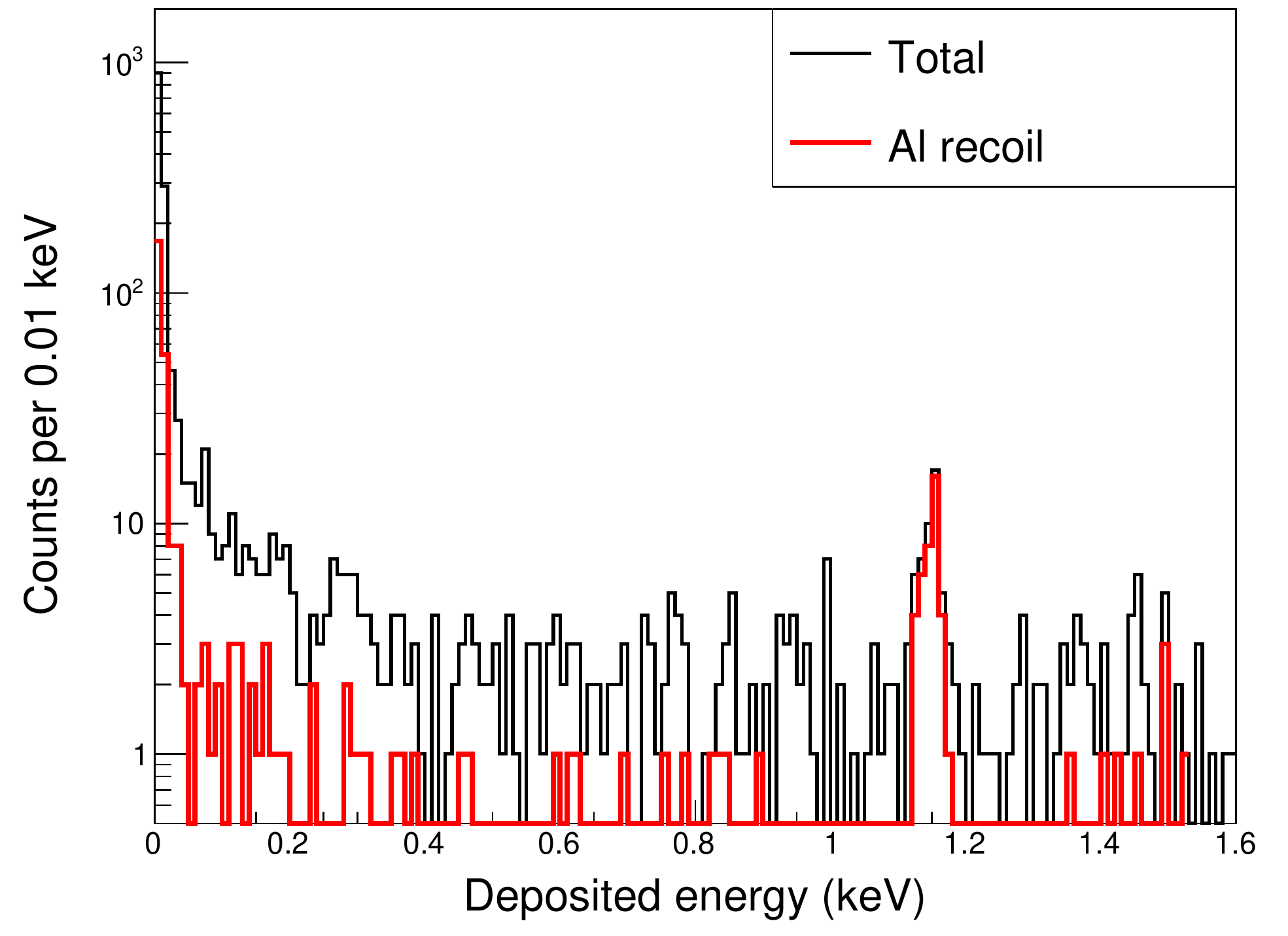}
	\caption{\label{fig:SappNCalSim}
    Geant4 simulation of the total energy (\emph{black} histogram) deposited in a \ce{Al_2O_3} crystal by \num{1.1e9} neutrons from an AmBe source under the condition of the CRESST n-calibration campaign described in the text. Probabilities for \textgamma{} emission are taken from tabulated FIFRELIN/Iradina simulations \cite{f4g4data}. Besides continuous contributions from electron recoils and oxygen recoils, the simulation predicts a clear peak at \SI{1144}{\eV} caused purely by aluminium recoils (\emph{red} histogram). The simulated spectrum is convolved with the empirical energy resolution of the studied detector that is $\sigma=\SI{12.3}{\eV}$ at the peak position.
    }
\end{figure}

We used ImpCRESST \cite{Abdelhameed2019}, based on Geant4 \cite{GEANT4:2002zbu,Allison:2006ve,Allison:2016lfl} version 10.6.3, to perform a Monte Carlo simulation of the AmBe neutron calibration: emission of the neutrons at the precise source position outside the lead and copper shieldings of the CRESST setup, scattering, and thermalisation of the neutrons in an accurate geometry of the setup and cryostat, and finally capturing of the neutrons on \ce{^{27}Al} nuclei of the \ce{Al_2O_3} absorber crystals of the CRESST detector. To simulate the nuclear recoils caused by \ce{^{28}Al} as primary knock-on atom (PKA) after the $\ce{^{27}Al}(\mathrm{n},\gamma)\ce{^{28}Al}$ capture process, an accurate model of the nuclear de-excitation process is needed, especially accurate probabilities for single-$\gamma$ and multi-$\gamma$ emissions. As standard Geant4 is lacking those \cite{Thulliez_2021}, we now enhanced ImpCRESST compared to a similar, earlier simulation reported in Ref.~\cite{PhysRevD.108.022005,fuss2022}: we are using the fifrelin4geant4 library \cite{f4g4} to read in tabulated emission probabilities \cite{f4g4data} of the \ce{^{28}Al} de-excitation cascade calculated \cite{PhysRevD.108.072009} with the codes FIFRELIN \cite{Litaize2015} and Iradina \cite{BORSCHEL20112133}. 
Simulating in total \num{1.1e9} neutrons started at the position of the AmBe source, equivalent to \SI{6.4}{\day} of calibration,
we found the total energy deposition in the \ce{Al_2O_3} crystal as shown in Fig.~\ref{fig:SappNCalSim} (\emph{black} histogram).

It considers the detector response function that consists of a convolution with the empirical energy resolution of the detector, which is based on the reconstructed amplitudes of simulated pulses
\begin{equation}
    \sigma(E)=\sqrt{\sigma_\mathrm{thr}^2+\SI{93}{\meV}\cdot(E-E_\mathrm{thr})}
\end{equation}
and a finite time resolution of \SI{2}{\ms} over which the simulated energy depositions are accumulated. Even with the applied energy resolution, a clear peak at \SI{1144}{\eV} is visible that is purely caused by recoiling Al nuclei (\emph{red} histogram) in agreement with the expectation.

\begin{figure}[h]
\centering
\subfloat[Subfigure 1][]{
\includegraphics[width=0.48\textwidth]{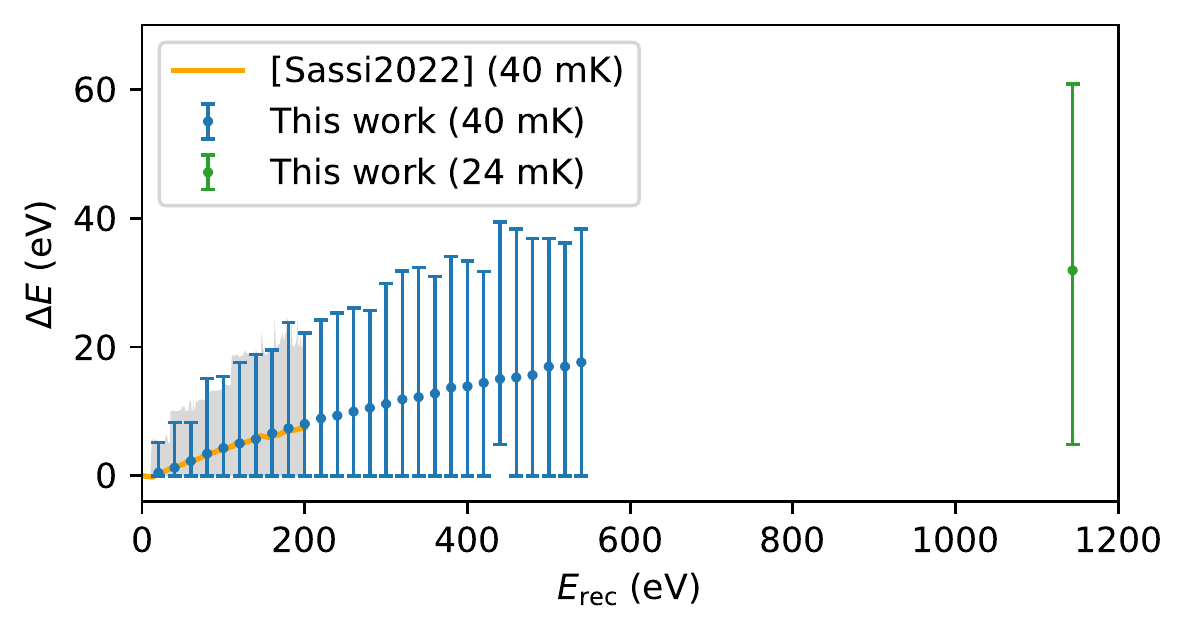}
\label{fig:DefectEnergy}}\\
\subfloat[Subfigure 2][]{
\includegraphics[width=0.46\textwidth]{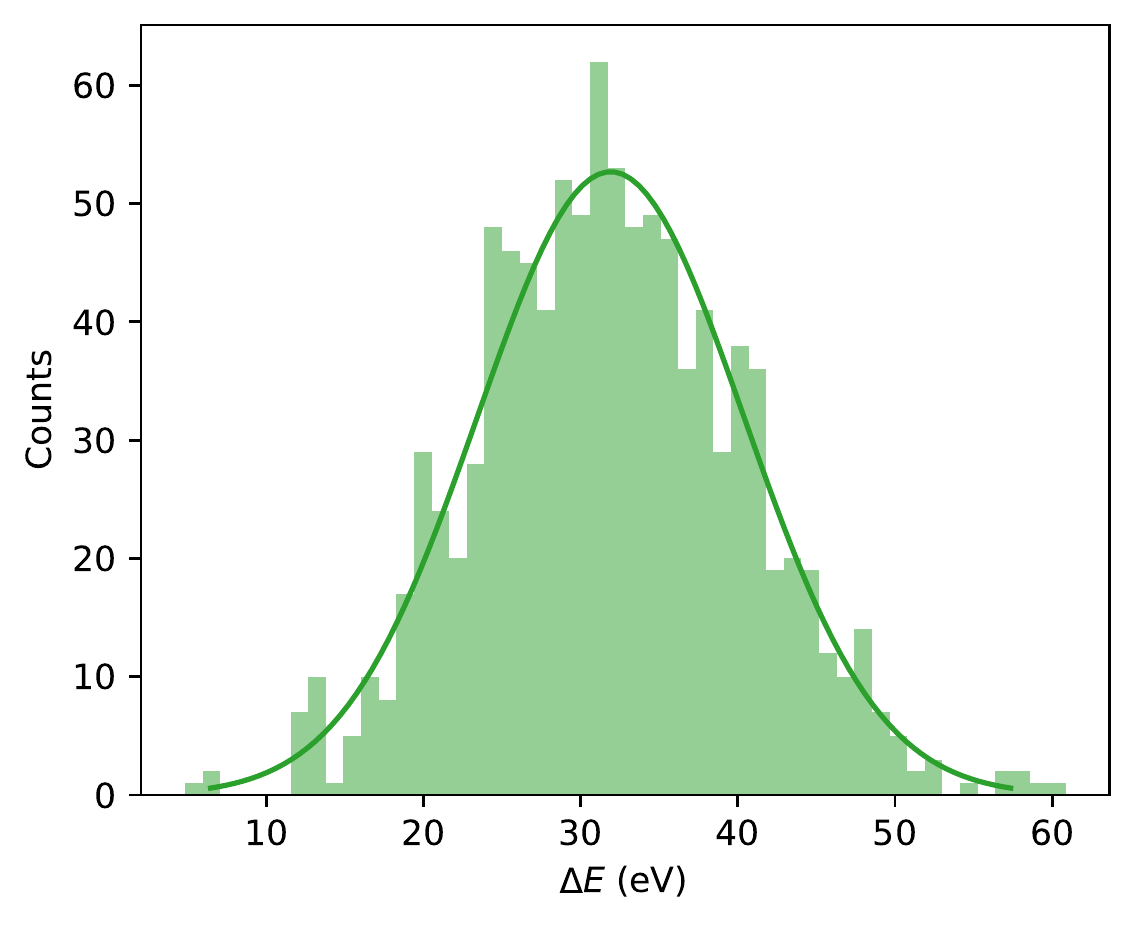}
\label{fig:DefectEffect}}
\caption{LAMMPS simulation of the energy loss $\Delta E$ due to the creation of crystal defects by a recoiling Al atom of kinetic energy $E_\mathrm{rec}$ as primary knock-on atom: \protect\subref{fig:DefectEnergy} for each $E_\mathrm{rec}$ value, \num{1000} random recoil directions and start positions were simulated. The \emph{data points} give the resulting mean value of $\Delta E$, the error bars give the minimal and maximal values. To allow comparability with the results of \citeauthor{PhysRevD.106.063012} \cite{PhysRevD.106.063012} (\emph{yellow  line} and \emph{gray histogram}) and to extend it, recoils at $E_\mathrm{rec} < \SI{540}{\eV}$ are simulated at \SI{40}{\milli\kelvin} (\emph{blue} data points). The recoil at \SI{1144}{\eV} is simulated at \SI{24}{\milli\kelvin} (\emph{green} data point), the actual operation temperature of the CRESST detector. For this latter data point, \protect\subref{fig:DefectEffect} shows the underlying $\Delta E$ distribution. It can be fitted by a Gaussian $G(\Delta E; \mu, \sigma)$ with a mean of $\mu=\SI{31.9}{\eV}$ and a standard deviation of $\sigma=\SI{8.5}{eV}$.}
\label{fig:CrystalDefects}
\end{figure}

To consider the effect of defect creation inside the \ce{Al_2O_3} absorber crystal, we followed the approach of \citeauthor{PhysRevD.106.063012} \cite{PhysRevD.106.063012} and performed a Molecular Dynamics (MD) simulation with the LAMMPS software~\cite{Thompson2022}. We used the interatomic potential of \citeauthor{Vashishta2008} \cite{Vashishta2008} and the unit cell definition given in Ref.~\cite{Lewis:a20727}. Using a supercell of $21 \times 21 \times 9$ times the unit cell ensures that an Al-PKA with a kinetic energy of $E_\mathrm{rec} = \SI{1144}{\eV}$ is completely contained before it reaches the boundary. Periodic boundary conditions were applied to the supercell. The energy loss $\Delta E$ to the creation of crystal defects is determined as the difference in potential energy of the crystal before the recoil and \SI{8}{\ps} afterwards. We validate our simulation by reproducing the results of \citeauthor{PhysRevD.106.063012} at \SI{40}{\milli\kelvin} and $E_\mathrm{rec} < \SI{200}{\eV}$, see \emph{yellow, gray} vs.\ \emph{blue} data in Fig.~\ref{fig:DefectEnergy}. For each $E_\mathrm{rec}$ value, we randomly sampled \num{1000} recoil directions and start positions within the central unit cell. Afterwards, we simulate the \SI{1144}{\eV}-PKA at the actual temperature of the CRESST absorber crystal of \SI{24}{\milli\kelvin}, see \emph{green} data in Fig.~\ref{fig:CrystalDefects}. \Cref{fig:DefectEffect} shows the obtained energy loss distribution which can be described by a Gaussian $G(\Delta E; \mu, \sigma)$ with a mean of $\mu=\SI{31.9}{\eV}$ and a standard deviation of $\sigma=\SI{8.5}{eV}$. We found that considering quantum fluctuations in LAMMPS via the quantum thermal bath (qtb) scheme  \cite{Dammak2009,Barrat2011} does not significantly affect these results. Details of the MD simulation will be given in a forthcoming paper.

\section{Data analysis} \label{Analysis}
In this section we briefly describe the most important steps in our analysis. A detailed description of the analysis of this detector can be found in Ref.~\cite{fuchs2023}.

\subsection{Data processing and calibration}
We use a dead-time free continuous data acquisition to save the detector output to disk \cite{ferreiro2018}. The data are then triggered using an optimum filter \cite{GattiManfredi1986} at a threshold corresponding to an accepted number of one noise trigger per kg$\cdot$d of exposure. A detailed description of the method of trigger threshold determination can be found in Ref.~\cite{MANCUSO2019492}. The detector presented in this work was triggered at a threshold corresponding to $E_{\mathrm{thr}} = 52\,$eV with a resolution of $\sigma_{\mathrm{thr}} = 7\,$eV at threshold. The optimum filter is constructed by incorporating the known signal shape and noise power spectrum of the detector. We obtain the signal shape by fitting a parametric pulse model \cite{Pröbst1995} to an averaged pulse from a list of cleaned nuclear recoil events from the linear regime of the detector. The noise power spectrum is created from the ensemble average of a list of cleaned empty noise traces. Cleaned in this context means the removal of any remaining pulses from the random selection of traces from the continuous stream. 

We apply a list of selection criteria to the triggered pulses which are designed to keep only pulses for which we can properly reconstruct the amplitude. The amplitude reconstruction is performed via the optimum filter during the triggering of the data.

The energy calibration is done using the K$_{\alpha}$ and K$_{\beta}$ lines of a low-activity $^{55}$Fe source with an average energy of 5.89$\,$keV and 6.49$\,$keV, respectively \cite{international2007iaea}. The source is mounted in direct vicinity of the absorber crystal, covered by a layer of glue. In order to prevent Auger electrons or scintillation photons of the glue to reach the detector, the source is coated with a thin layer of gold. The response of the detector to events caused by particle interactions is linearized using the response of the injected heater pulses.

\subsection{Signal efficiency} \label{Eff}
\begin{figure}[h]
\centering
\includegraphics[width=0.5\textwidth]{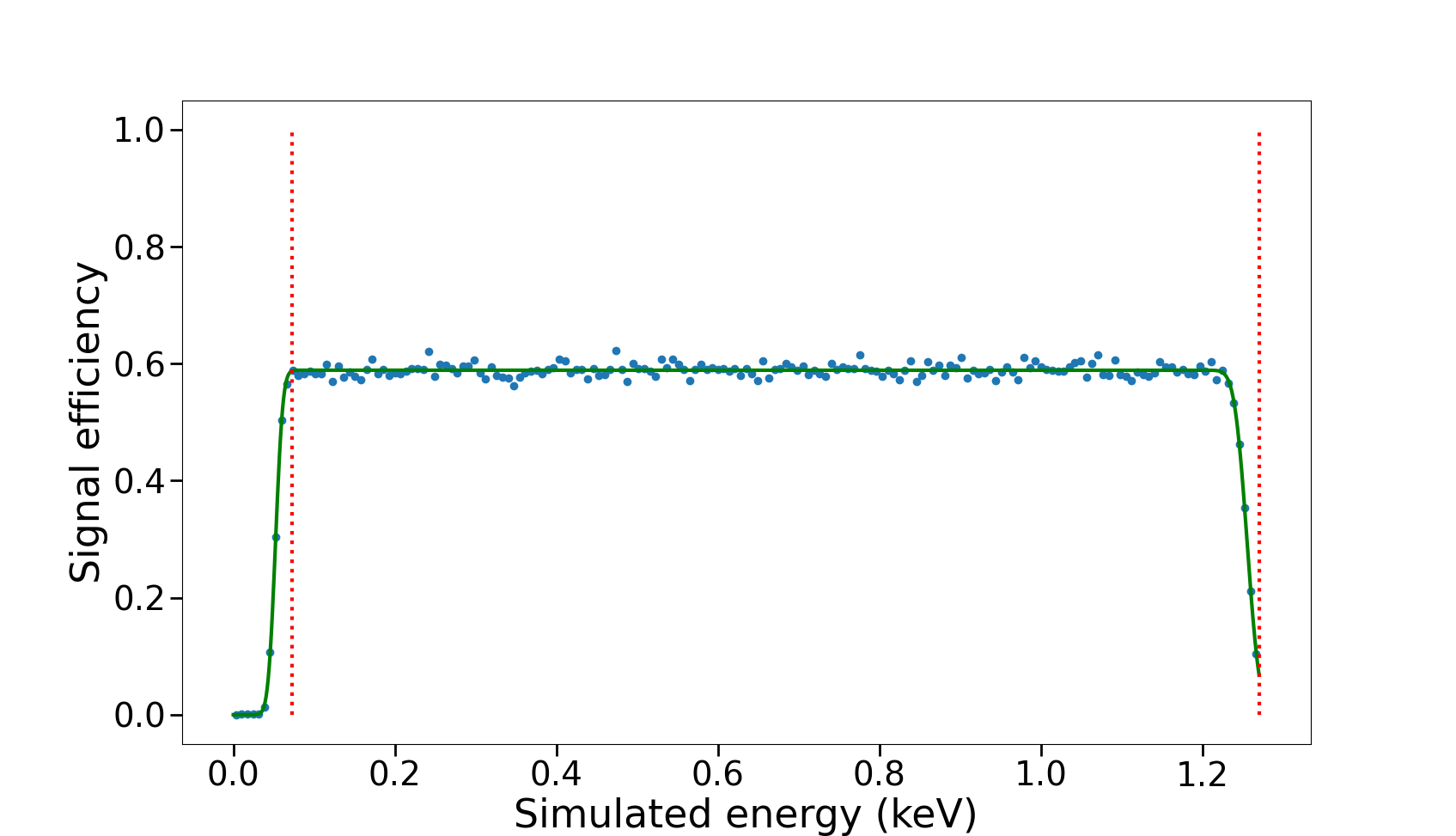}
\caption{\label{fig:Eff} Energy dependent signal efficiency of simulated events after triggering and applying all selection criteria. The blue dots show the output of our simulation. The green line shows a fit of the efficiency, which is used to rescale our fit model. The red dotted line show the energy region of interest (ROI) included in the analysis.}
\end{figure}

We perform a simulation to obtain the probability of valid events to survive the triggering and our selection criteria by superimposing artificial pulses, scaled to different energies and randomly distributed in time, onto the raw data stream. We then determine the surviving fraction of simulated pulses after passing through the same analysis chain as the real data. The energy dependent signal efficiency of events is shown in Fig.~\ref{fig:Eff}.

We fit the energy dependent signal efficiency $\varepsilon(E)$ with two error functions:

\begin{equation} \label{eq:EffFit}
    \varepsilon(E) = \frac{1-p_{1}}{2} \cdot  \left( erf(\frac{E - E_{\mathrm{thr}}}{\sqrt{2}\sigma_{\mathrm{thr}}}) - erf(\frac{E - E_{\mathrm{lin}}}{\sqrt{2}\sigma_{\mathrm{lin}}}) \right) + p_{2}
\end{equation}

with $1 - p_{1} = 0.59$ describing the flat plateau over most of the energy range. Due to the deadtime of the trigger, mostly caused by the presence of injected heater pulses that are used in the calibration, the trigger efficiency never reaches 100$\,\%$. The energy independent selection criteria further reduce the signal efficiency to about 59$\,\%$. The parameter $p_{2} = 10^{-4}$ describes the very small probability of events far below the energy threshold surviving the trigger. This is mostly due to a random coincidence of a small simulated event with a larger real event. These events are efficiently removed from the simulation. The parameters of the first error function reflect the threshold $E_{\mathrm{thr}} = 52\,$eV and the resolution at threshold $\sigma_{\mathrm{thr}} = 7\,$eV. The second error function describes the upper edge of our region of interest (ROI). We perform a cut in the data at a reconstructed amplitude corresponding to the transition into the non-linear regime of the TES, which lies well below the point at which pulses start to saturate. This is a very conservative analysis choice. Since we use the information of the injected heater pulses to linearize the response of the detector up to higher energies, this cut is not strictly necessary. Equivalently to the effect at the threshold, both edges of the ROI have a sharp cutoff in units of reconstructed amplitudes, which get smeared by the energy resolution in units of energy, leading to the error functions. The fit parameters for the second error function are $E_{\mathrm{lin}} = 1.255\,$keV and $\sigma_{\mathrm{lin}} = 12.3\,$eV.

We use this fit function to rescale our fit model within our ROI. As the lower limit we choose the energy of the trigger threshold plus three times the energy resolution $E_{\mathrm{low}} = E_\mathrm{thr}+3 \cdot \sigma_\mathrm{thr} = 73\,$eV. The choice for the upper edge is more critical regarding the search for a peak at 1.144$\,$keV and is therefore set to $E_{\mathrm{up}} = 1.27\,$keV.

\subsection{Energy spectrum and empirical fit function}

\begin{figure}[h]
\centering
\includegraphics[width=0.5\textwidth]{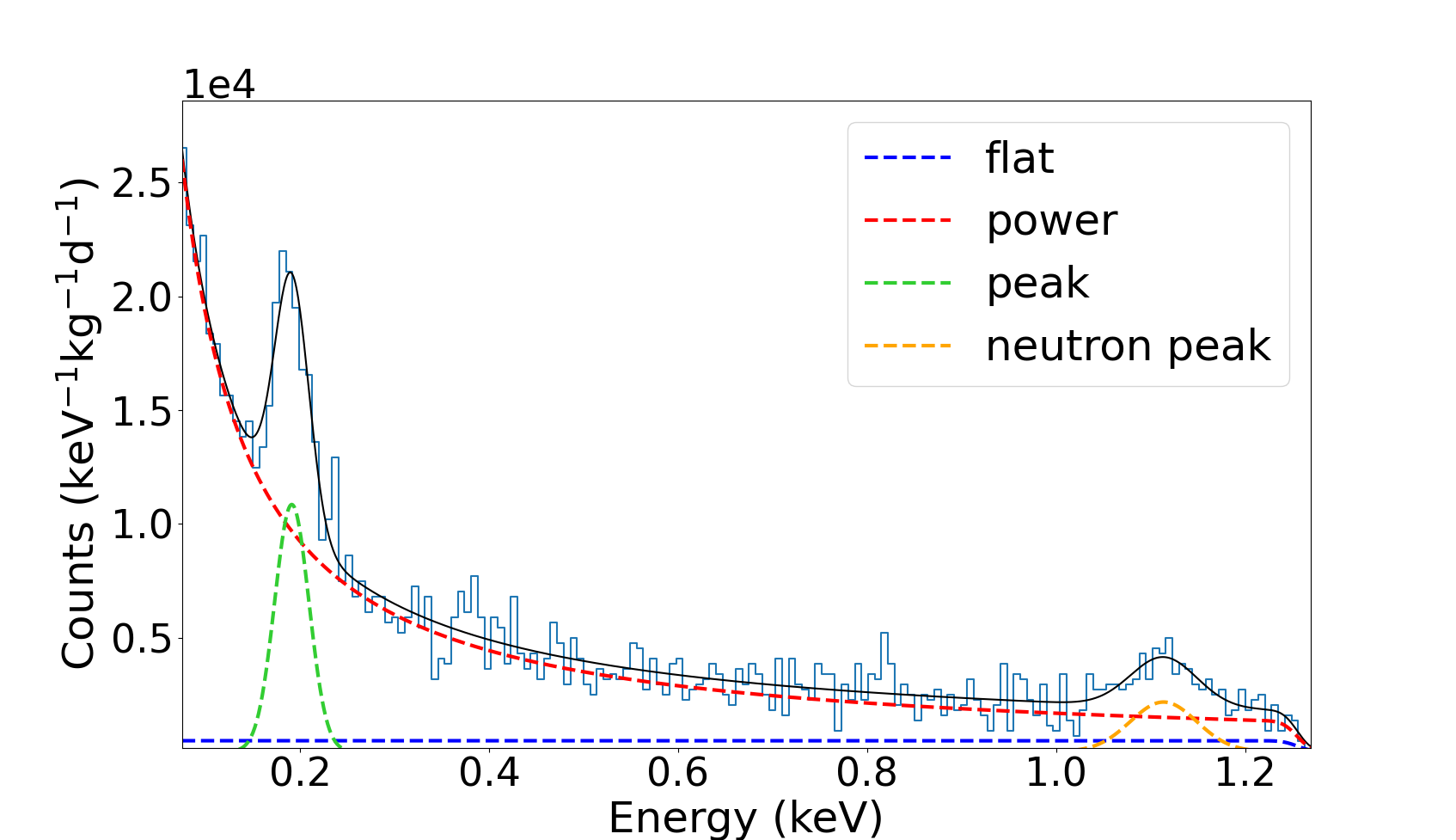}
\caption{\label{fig:NCalFits} Fit of the measured spectrum, corrected by the energy dependent survival probability of events and scaled by the exposure. The fit contains a Gaussian component to account for the feature of unknown origin at around 190 eV and a flat component. The strong rise of the event rate at low energies is modeled by a power law. An additional Gaussian distribution is describing the potential neutron peak at around 1.1$\,$keV.}
\end{figure}

The measured spectrum of the neutron calibration is scaled by the exposure of 0.634$\cdot$kg$\,$d and binned in 7$\,$eV bins. The resulting histogram can be seen in Fig.~\ref{fig:NCalFits}.

The strongly rising event rate towards lower energies is empirically described by a power law, $A \cdot E^{-p}$, with amplitude $A$ (number of events) and exponent $p$. Additionally, we fit a constant component, $C$ (number of events), to the whole energy spectrum. The most prominent feature in this spectrum is a peak at roughly 190$\,$eV, whose origin is not yet fully understood. This peak appears in the energy spectrum of this detector independently of the presence of the AmBe source. A more detailed analysis in Ref.~\cite{fuchs2023} excludes Compton scattering of $^{55}$Fe events or X-rays of surrounding materials as possible sources. The rate of events in this peak decreases with a decay time compatible with the half-life of $^{55}$Fe. Therefore, we consider scintillation light originating from the glue covering the source as a possible explanation. In the following we refer to this peak at 190$\,$eV as \textit{unknown feature}, in order to distinguish it from the sought for nuclear recoil peak at 1.1$\,$keV. Both, the feature of unknown origin at around 190 eV, as well as the peak at around 1.1 keV are fit with a Gaussian distribution. The full function is given by: 

\begin{equation} \label{eq:Fitfunc}
\begin{split}
	f(E) = & \varepsilon(E) \cdot \left(\frac{A \cdot E^{-p}}{\mathcal{N}_{\mathrm{A}}}  + \dfrac{C}{\mathcal{N}_{\mathrm{C}}} \right.  \\ 
    & +  \left. \sum_{i = \mathrm{F,P}} \frac{N_{i} \cdot \mathcal{G}(E; \mu_{i}, \sigma_{i})}{\mathcal{N}_{i}} \right) \cdot \Theta(E \in \mathrm{ROI})
\end{split}
\end{equation}

where $\varepsilon(E)$ is the energy dependent signal efficiency, $N_{i}$, $\mu_{i}$ and $\sigma_{i}$ are the amplitude (number of events), mean and standard deviation of the two Gaussian distributions of the unknown feature ($\mathrm{F}$) and the peak ($\mathrm{P}$). All components are only fit using data within our chosen ROI and each component has a normalization factor $\mathcal{N}_{x}$, which is the integral of the respective component over the ROI.

\subsection{Likelihood framework}

We estimate the parameters by maximizing a Poissonian likelihood function: 

\begin{equation} \label{eq:BinnedL}
	\mathcal{L} = \prod\limits_{i=0}^{m} e^{-\Tilde{f_{i}}} \cdot \dfrac{(\Tilde{f_{i}})^{k_{i}}}{k_{i}!}
\end{equation}

with $m$ being the number of bins and $k_{i}$ being the observed number of events in the $i$th bin. $\Tilde{f_{i}}$ is the expected number of events in the $i$th bin based on the fit function of Eq.~\ref{eq:Fitfunc}, corrected by the efficiency in that bin, $\varepsilon_{i}$, such that:

\begin{equation}
    \sum_{i = 0}^{m} \frac{\Tilde{f_{i}}}{\varepsilon_{i}} = A + C + N_{\mathrm{F}} + N_{\mathrm{P}}
\end{equation}

We calculate the statistical significance of the peak using a test statistic defined by a likelihood ratio:

\begin{equation} \label{eq:PeakTS}
    q_{\mathrm{peak}} = -2 \cdot ln \left( \dfrac{\mathcal{L}(N_{\mathrm{P}} = 0,\, \hat{\hat{\boldsymbol{\theta}}})}{\mathcal{L}(\hat{N}_{\mathrm{P}},\, \hat{\mu}_{\mathrm{P}},\, \hat{\sigma}_{\mathrm{p}},\, \hat{\boldsymbol{\theta}})} \right)
\end{equation}

In the denominator we estimate the best fit (denoted by a hat) of the amplitude, standard deviation and position of the signal peak ($\hat{N}_{\mathrm{P}},\, \hat{\sigma}_{\mathrm{P}},\, \hat{\mu}_{\mathrm{P}}$) as well as all other (nuisance) parameters, $\hat{\boldsymbol{\theta}}$. The numerator maximizes the likelihood under the condition of no additional signal peak ($N_{\mathrm{P}} = 0$). The best estimator of the parameters under the conditional likelihood are denoted by a double hat. \\

For the investigation of a potential shift of the peak due to energy loss to defect creation we need to take the uncertainty of our calibration into account. Therefore we introduce an additional parameter $c_{\mathrm{E}}$, acting as a multiplicative factor. This parameter is constrained by a Gaussian term centered around one and with a standard deviation equal to the uncertainty on the energy calibration $\Delta c_{\mathrm{E}}$. The likelihood function is transformed accordingly:

\begin{equation} \label{eq:Likelihood_cpe}
\mathcal{L_{\mathrm{E}}}(\mu_{\mathrm{P}}, \boldsymbol{\theta}, \mathrm{E}) = \mathcal{L}(\mu_{\mathrm{P}}, \boldsymbol{\theta}, c_{\mathrm{E}} \cdot \mathrm{E}) \times \mathcal{G}(c_{\mathrm{E}}; \mu = 1, \sigma = \Delta c_{\mathrm{E}})
\end{equation}

with $\mu_{\mathrm{P}}$ being the peak position and $\boldsymbol{\theta}$ being all other parameters. The fit can shift the energy spectrum by the factor $c_{\mathrm{E}}$, which is constrained by the uncertainty $\Delta c_{\mathrm{E}}$.

The corresponding test statistic for the calculation of the significance of a measured shift is:

\begin{equation} \label{eq:ShiftTS}
q_{\mathrm{shift}} = \begin{cases}
-2 \cdot ln \, \lambda(\mu_{\mathrm{P}} = 1144\,\mathrm{eV}) &\text{, $\hat{\mu} < 1144$\,$\mathrm{eV}$}\\~\\
0 &\text{, $\hat{\mu} \geq 1144$\,$\mathrm{eV}$}
\end{cases}
\end{equation}

with:

\begin{equation}
    \lambda(\mu_{\mathrm{P}} = 1144\,\mathrm{eV}) = \dfrac{\mathcal{L_{\mathrm{E}}}(\mu_{\mathrm{P}} = 1144\,\mathrm{eV}, \hat{\hat{\boldsymbol{\theta}}}, \hat{\hat{c_{\mathrm{E}}}} \cdot \mathrm{E})}{\mathcal{L_{\mathrm{E}}}(\hat{\mu}_{\mathrm{p}}, \hat{\boldsymbol{\theta}}, \hat{c_{\mathrm{E}}} \cdot \mathrm{E}) }
\end{equation}

In the denominator we estimate the parameters that maximize the likelihood (denoted by a hat). In the numerator we estimate parameters that maximize the likelihood under the condition that the position of the peak is fixed at $\mu_{\mathrm{P}} = 1144\,$eV (denoted by a double hat), which corresponds to the expected position without energy loss (see Tab.~\ref{tab:Al27_properties}).

\subsection{Uncertainty estimation} \label{UncertRange}

In order to be able to accurately determine the statistical significance of a shift in energy, we need an estimate on the uncertainty of our calibration, $\Delta c_{\mathrm{E}}$. For the calibration we use a combination of a mono-energetic calibration source ($^{55}$Fe) and injected electric heater pulses of different energies, covering the entire dynamic range of the TES. The reconstructed amplitudes of the heater pulses are used to linearize the detector response. The calibration of the detector then relies on the linear extrapolation from the energy of the $^{55}$Fe calibration source (5.89$\,$keV, 6.49$\,$keV) down to the threshold, which is typically at $\mathcal{O}$(10)$\,$eV.

The exact uncertainty on our calibration is difficult to assess, but we identify two main sources of uncertainty in our calibration:

\begin{itemize}
    \item The precision on the reconstructed position of the calibration peak in the energy spectrum
    \item The method of linearization of the detector response
\end{itemize}

\paragraph{Fit of the calibration peak:} We fit two Gaussian distributions to the peaks of the $^{55}$Fe source in the spectrum of reconstructed and linearized amplitudes of events. We then take the uncertainty on the resulting peak position as an estimator for the uncertainty of our calibration, which results in $\Delta c_{\mathrm{E}} = 0.056\,\%$.

\paragraph{Correction of the time dependence:} The linearization is done via two separate spline interpolations. In the first step we compute the time dependent detector response for each of the injected heater pulse amplitudes via Gaussian kernel smoothing. The heater pulses are injected periodically every few seconds. The default choice of the kernel size for the smoothing is 30$\,$min. The effect of varying the kernel size in an interval between 3$\,$min and 60$\,$min on the reconstructed peak position is less than 0.06$\,\%$.

\paragraph{Linearization of the detector response:} In a second step, we convert the amplitudes of reconstructed particle events in the detector into heater pulse equivalent energy by evaluating the time dependent detector response at the exact timestamp of each particle interaction with a second spline interpolation. For this spline interpolation we fix the lowest point at zero. The impact of leaving the lowest point as a free floating offset on the reconstructed calibration peak position is negligibly small.

Another possible source of a systematic uncertainty can be a difference in the response of the detector between particle events and injected heater pulses, in particular in the non-linear regime of the TES, as suggested by the authors of Ref.~\cite{abele2025subkevelectronrecoilcalibration}.

Nevertheless, we compute the statistical significance of an observed shift for a range of different uncertainties, ranging from our estimated value of $\Delta c_{\mathrm{E}}$ = 0.056$\,\%$ up to a more conservative, but arbitrary value of 2$\,\%$.

We also simulate Monte Carlo (MC) data at each uncertainty under the null hypothesis (no shift, $\mu_{\mathrm{P}} = 1144\,$eV) and under the alternative hypothesis (shift by 32$\,$eV due to energy loss, $\mu_{\mathrm{P}} = 1112\,$eV) in order to estimate the expected distributions of the test statistic under both hypotheses and compare them to the observed value. In both cases we draw a random value $c_{E, \mathrm{rnd}}$ from \mbox{$\mathcal{G}(c_{\mathrm{E}}; \mu = 1, \sigma = \Delta c_{\mathrm{E}})$} for each individual simulation. 

\section{Results} \label{Results}

The resulting parameters of the best fit of Eq.~\ref{eq:Fitfunc} are listed in Tab.~\ref{tab:NCal_FitResults} toghether with their 1$\,\sigma$ confidence intervals. The different components of the fit function are also shown together with the energy spectrum in Fig.~\ref{fig:NCalFits}

\begin{table}[!htb]
\centering
	\caption{\label{tab:NCal_FitResults}Resulting parameters and their 1$\,\sigma$ confidence intervals of an empirical fit of the neutron calibration spectrum with Eq.~\ref{eq:Fitfunc}.}
	\newcolumntype{C}{>{\centering\arraybackslash}X}
	\setlength\extrarowheight{3pt}
	\noindent
	\begin{tabularx}{0.48\textwidth}{m{1em}  l C}
	\hline          
        \multicolumn{2}{c}{Parameter} & Fit result\\ \hline
        \multicolumn{3}{l}{Constant component}\\
        & Amplitude $C$ (events) & 609$^{+443}_{-500}$  \\
        \multicolumn{3}{l}{Exponential component}\\
        & Amplitude $A$ (events) & 5514$^{+505}_{-452}$ \\
        & Exponent $p$ & 1.058$^{+0.074}_{-0.073}$  \\
        \multicolumn{3}{l}{Gaussian feature at \SI{190}{\eV}}\\
        & Amplitude $N_{\mathrm{F}}$ (events) & 528$^{+64}_{-60}$  \\
        & Mean $\mu_{\mathrm{F}}$ (eV) & 191.3$^{+2.2}_{-2.1}$  \\
        & Standard deviation $\sigma_{\mathrm{F}}$ (eV) & 18.1$^{+2.6}_{-2.3}$  \\
        \multicolumn{3}{l}{Gaussian peak at \SI{1.1}{\keV}}\\
        & Amplitude $N_{\mathrm{P}}$ (events) & 205$^{+42}_{-39}$ \\
        &Mean $\mu_{\mathrm{P}}$ (eV) & 1113.6$^{+6.5}_{-6.5}$  \\
        & Standard deviation $\sigma_{\mathrm{P}}$ (eV) & 35.1$^{+7.1}_{-6.5}$  \\ \hline
      \end{tabularx}
\end{table}

The observed test statistic of Eq.~\ref{eq:PeakTS} is $q_{\textrm{peak}} = 48.86$. We simulate 3.3 million MC datasets containing no peak and compute the corresponding distribution of $q_{\textrm{peak}}$, without any constraints on the fit. The distribution deviates from the expected shape of a $\chi^{2}_{3}$ distribution, showing a much steeper tail towards higher values. Based on the very conservative approximation with a $\chi^{2}_{3}$ distribution, we calculate a two-sided significance of 6.4$\,\sigma$.

\begin{figure}[h]
\centering
\includegraphics[width=0.5\textwidth]{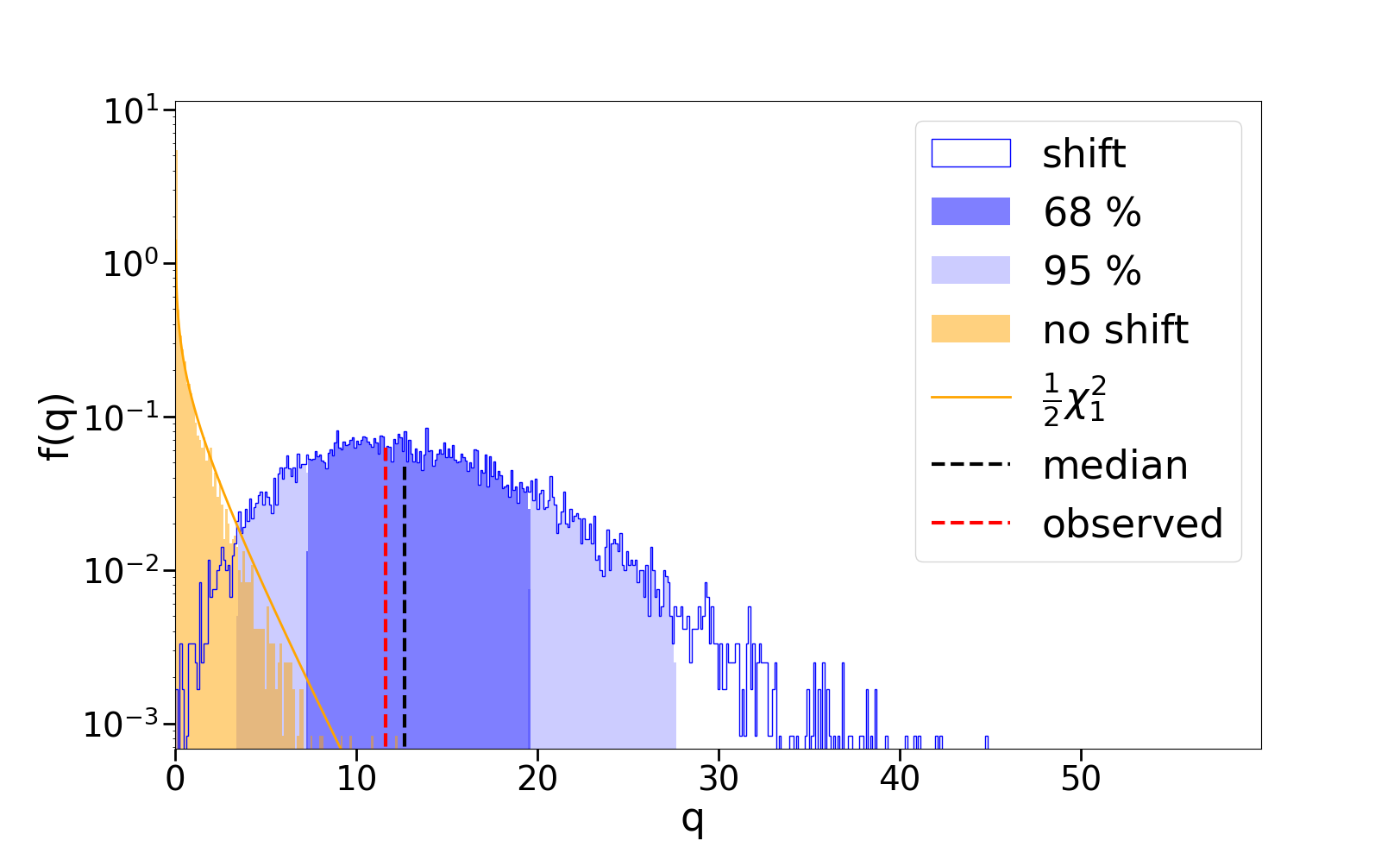}
\caption{\label{fig:shiftTS_example} Example of the distribution of the test statistic $q_{\mathrm{shift}}$ assuming an uncertainty of $\Delta c_{\mathrm{E}}= 0.5\,\%$ on the energy calibration. In orange: The distribution under the null hypothesis (no shift in the simulated data) can be described by a half-$\chi^{2}_{1}$ distribution. In blue: The distribution of the test statistic under the alternative hypothesis (shift of 32$\,$eV due to defect creation). The blue shaded regions show the ranges containing 68$\,\%$ and 95$\,\%$ of the values. The red dashed line shows the observed value of the test statistic.}
\end{figure}

The extracted position of the peak at (1113.6$^{+6.5}_{-6.5}$)$\,$eV shows a clear deviation from the expected value of 1144$\,$eV (see Tab.~\ref{tab:Al27_properties}), and is compatible with the shift expected from our simulations, shown in Fig.~\ref{fig:DefectEffect}. Whether this deviation can be explained by an energy loss due to crystal defect creation or by a statistical fluctuation depends strongly on the uncertainty of our energy calibration. Fig.~\ref{fig:shiftTS_example} shows the distribution of the test statistic $q_{\mathrm{shift}}$ of Eq.~\ref{eq:ShiftTS} under both hypotheses. An uncertainty of $\Delta c_{\mathrm{E}} = 0.5\,\%$ is chosen as an example. We repeat the same comparison for the full range of uncertainties defined in Sec.~\ref{UncertRange} from 0.056$\,\%$ up to 2$\,\%$.

Shown in orange in Fig.~\ref{fig:shiftTS_example} is the distribution of the test statistic $q_{\mathrm{shift}}$ under the null hypothesis (no shift of the nuclear recoil peak in the simulated data). It can be described by a half-$\chi^{2}_{1}$ distribution, which is used to translate the observed p-value into a one-sided statistical significance in units of Gaussian standard deviations $\sigma$. We observe slight deviations from this behaviour for smaller uncertainties, where the tail of the true distribution is falling steeper than in the half-$\chi^{2}_{1}$ approximation. This effect is already slightly visible in the example in Fig.~\ref{fig:shiftTS_example} and gets more pronounced toward smaller uncertainties. This suggests that we underestimate the significance of the shift for uncertainties smaller than $\sim$0.5$\,\%$.
The blue histogram shows the distribution of the test statistic when the peak in the simulated data is shifted by 32$\,$eV, corresponding to the expected energy loss due to defect creation in the crystal. Lastly, the red line shows the observed value of the test statistic.

\begin{figure}[h]
\centering
\includegraphics[width=0.5\textwidth]{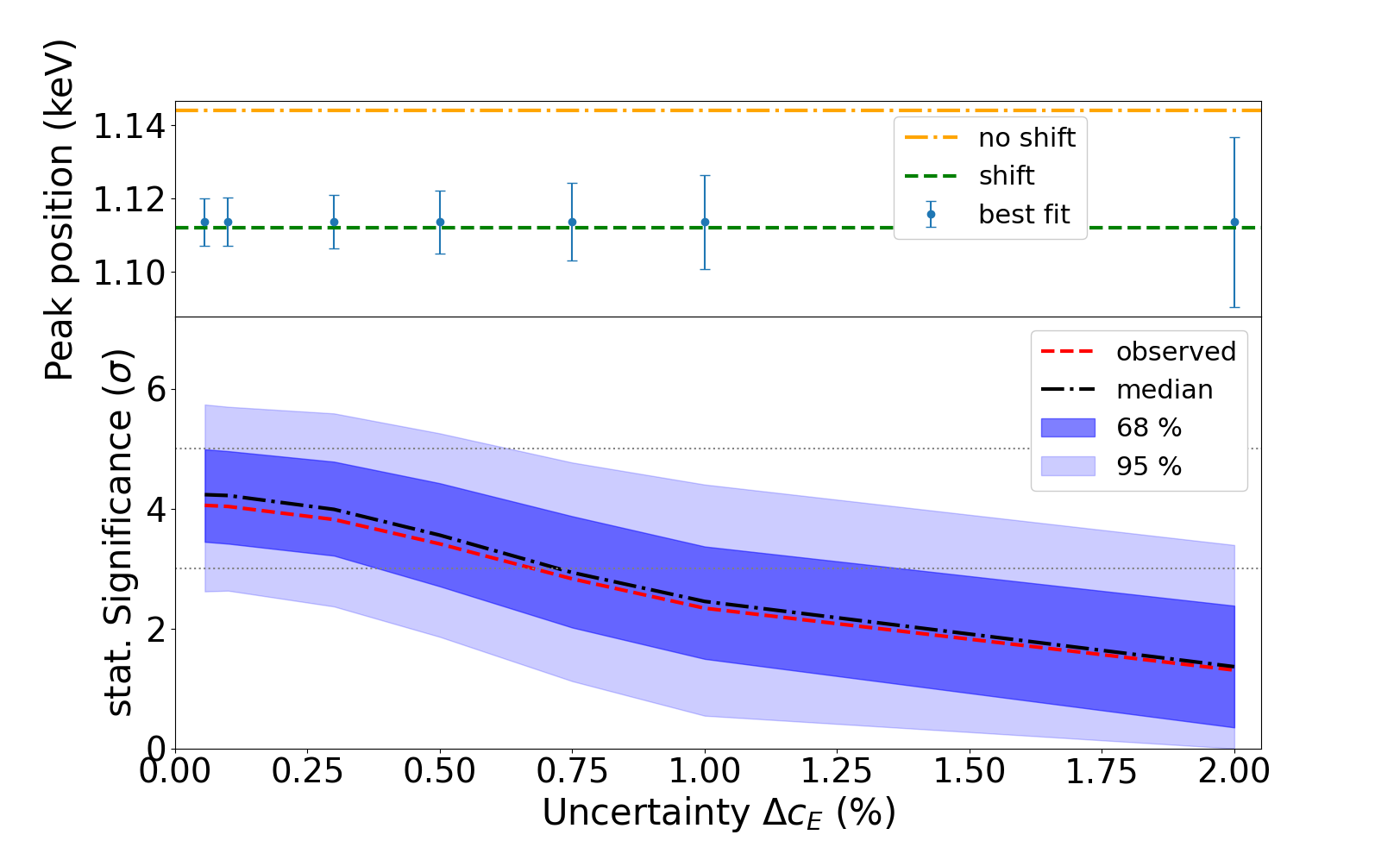}
\caption{\label{fig:sigVsUncert} Top: Best fit of the peak position and 1$\,\sigma$ confidence intervals as a function of the uncertainty on the calibration. Also shown is the expected result in case of no shift (dashdotted orange line) and in case of a shift of 32$\,$eV due to energy loss to defect creation (green dashed line). Bottom: Statistical significance of the observed shift as a function of the uncertainty on the calibration shown as the red dashed line. The blue bands show the expected distribution of the significance under the hypothesis of a shift of the peak position due to energy loss to defect creation, with the dark blue and the light blue region containing 68$\,\%$ and 95$\,\%$ of the values, respectively.}
\end{figure}

We calculate these distributions for different uncertainties $\Delta c_{\mathrm{E}}$ ranging from 0.056$\,\%$ to 2$\,\%$. The resulting best fit of the peak position as well as the statistical significance of the observed shift is shown as a function of the assumed uncertainty of our calibration in Fig.~\ref{fig:sigVsUncert}.

Independently of the uncertainty of the calibration, the best fit value of the peak position is consistent with the expected shift of 32$\,$eV within the uncertainties. Due to the direct correlation of the peak position and the calibration factor, the 1$\,\sigma$ confidence interval of the reconstructed peak position increases with an increasing uncertainty on the calibration. Consequently, the statistical significance of the shift drops with an increasing uncertainty on the calibration. Considering the estimate of the uncertainty to the best of our knowledge of $\Delta c_{\mathrm{E}}\,=\,$0.056$\,\%$, the shift has a statistical significance of 4$\,\sigma$ and is therefore unlikely to be caused by a statistical fluctuation. For an uncertainty of $\Delta c_{\mathrm{E}}\,\lesssim \,$0.7$\,\%$, the significance is above 3$\,\sigma$. At an uncertainty of $\Delta c_{\mathrm{E}}\,=\,$1.3$\,\%$, the significance is at $\sim$2$\,\sigma$ and at the largest tested uncertainty of 2$\,\%$, the statistical significance of the shift drops to 1.3$\,\sigma$. Since we cannot make a definite quantitative statement about the uncertainty of our calibration, future precision measurements will be needed to confirm the observation of a shift in the expected peak position. Alternative methods to the injected electric heater pulses used for the linearization of the detector response will be particularly important. New methods involve e.g. an LED based calibration system or the usage of X-ray flourescence (XRF) \cite{abele2025subkevelectronrecoilcalibration}.

\section{Conclusion} \label{Conclusion}
We report on the observation of a nuclear recoil peak at (1113.6$^{+6.5}_{-6.5}$)$\,$eV in the data of an Al$_{2}$O$_{3}$ detector during irradiation with neutrons from an AmBe source in the CRESST-III setup at LNGS. This peak is caused by nuclear recoils induced by the emission of mono-energetic $\gamma$ photons, which follow after the capture of a thermal neutron on $^{27}$Al. While the first experimental proof of this effect was recently measured with CaWO$_{4}$ crystals (with a capture of thermal neutrons in $^{182}$W) and published by the CRAB and the NUCLEUS collaboration in Ref.~\cite{PhysRevLett.130.211802} and by the CRESST collaboration in Ref.~\cite{PhysRevD.108.022005}, this is the first time this effect is experimentally measured on a sapphire crystal.

Therefore, we show that this novel method is also suitable for an accurate low energy calibration of crystals containing $^{27}$Al.

Additionally, we reconstruct the position of the nuclear recoil peak with a shift of (30.4$\,\pm\,$6.5)$\,$eV below the expected energy. Since the energy calibration is based on a \textgamma{} source, this could be a hint at differences between electronic recoil and nuclear recoil events. Our simulations predict an energy loss of 32$\,$eV at the relevant energy scale of nuclear recoils due to defect creation in the crystals. Due to the inaccessibility of the exact value of the uncertainty of our energy calibration, we cannot make a conclusive statement about the statistical significance of this effect. Should this shift be indeed caused by an energy loss to the creation of defects in the crystal, this would not only be the first experimental proof of this effect, but also have strong implications about the expected shape of DM or CE\textnu NS nuclear recoil spectra \cite{kadribasic2020crystal, PhysRevD.106.063012, SoumSidikov2024}, and about possible origins of the LEE \cite{PhysRevD.106.083009, Nordlund2024}. This further underlines the need to replace the current calibration methods with a nuclear recoil based one. The importance of the implications of a possible shift in the position of mono-energetic nuclear recoil peaks calls for more precision measurements with various targets in the near future in order to validate this effect, which could be performed at the recently commissioned CRAB facility \cite{abele2025crabfacilitytuwien}.

\section*{acknowledgments}

We are grateful to Laboratori Nazionali del Gran Sasso-INFN for their generous support of CRESST. This work has been funded by the Deutsche Forschungsgemeinschaft (DFG, German Research Foundation) under Germany's Excellence Strategy-EXC 2094-390783311 and through the Sonderforschungsbereich (Collaborative Research Center) SFB1258 'Neutrinos and Dark Matter in Astro- and Particle Physics', by BMBF Grants No. 05A2023 and No. 05A23WO4 and by the Austrian science fund (FWF) http://dx.doi.org/10.55776/PAT1239524 and by http://dx.doi.org/10.55776/I5420. J. B. and H. K. were funded through the FWF project http://dx.doi.org/10.55776/P34778 ELOISE. The Bratislava group acknowledges a partial support provided by the Slovak Research and Development Agency (Project No. APVV-21-0377). J.B., H.K., M.M., and V.P. acknowledge funding by Austria's Agency for Education and Internationalisation (OeAD) through project No. CZ 13/2023 INCIDENCE. The computational results presented were partially obtained using the Max Planck Computing and Data Facility (MPCDF) and using the CLIP cluster.




\bibliography{refs.bib}

\end{document}